\documentclass[12pt,preprint]{aastex}

\shorttitle{CAUSAL BACKREACTION CONCORDANT ACCELERATION}
\shortauthors{BOCHNER}

\begin{document}

\title{COSMIC ACCELERATION AND A NEW CONCORDANCE FROM CAUSAL BACKREACTION}

\author{Brett Bochner}
\affil{Department of Physics and Astronomy, 
Hofstra University, Hempstead, NY 11549}
\email{brett\_bochner@alum.mit.edu, phybdb@hofstra.edu}

\begin{abstract}
A phenomenological formalism is presented in which the apparent 
acceleration of the universe is generated by large-scale 
structure formation, thus eliminating the magnitude and 
coincidence fine-tuning problems of the Cosmological Constant 
in the Concordance Model, as well as potential instability 
issues with dynamical Dark Energy. The observed acceleration 
results from the combined effect of innumerable local 
perturbations due to individually virializing systems, 
overlapping together in a smoothly-inhomogeneous adjustment 
of the FRW metric, in a process governed by the causal flow 
of inhomogeneity information outward from each clumped system. 
After explaining why arguments from the literature claiming 
to place restrictive limits upon backreaction are not applicable 
in a physically realistic cosmological analysis, we present a 
selection of simply-parameterized models which are capable of 
fitting the luminosity distance data from Type Ia supernovae 
essentially as well as the best-fit flat $\Lambda$CDM model, 
without resort to Dark Energy, any modification to gravity, 
or a local void. Simultaneously, these models can reproduce 
measured cosmological parameters such as the age of the 
universe, the matter density required for spatial flatness, 
the present-day deceleration parameter, and the angular scale 
of the Cosmic Microwave Background to within a reasonable 
proximity of their Concordance values. A potential observational 
signature for distinguishing this cosmological formalism from 
$\Lambda$CDM may be a cosmic jerk parameter significantly in 
excess of unity.
\end{abstract}

\keywords{cosmological parameters --- cosmology: theory --- 
          dark energy --- large-scale structure of universe}

\section{\label{SecIntroMotiv}INTRODUCTION AND MOTIVATIONS} 

Evidence from the past decade or so indicating that we live 
in a spatially flat \citep{WMAP7yrLikeliParams}, accelerating 
\citep{PerlAccel99,RiessAccel98} universe, yet one in which the 
density of clustering matter is significantly less than the 
critical density \citep[e.g.,][]{TurnerCaseOm0pt33}, has led 
cosmologists to a surprising conclusion: the dominant cosmic 
component appears to be ``Dark Energy'', a smoothly-distributed 
material that must possess negative pressure in order to be able 
to drive the acceleration \citep[e.g.,][]{KolbTurner}. 

The true nature of this Dark Energy remains very much in doubt, 
however. The simplest version, a Cosmological Constant ($\Lambda$), 
suffers from well-known aesthetic difficulties, including 
two different fine-tuning problems: one being the issue that 
$\rho _{\Lambda}$ is $\sim$$120$ orders of magnitude smaller 
than what would be expected from the Planck scale 
\citep[e.g.,][]{KolbTurner}; and the other being a 
``Coincidence Problem" \citep[e.g.,][]{ArkaniHamedCoinc} which 
questions why, given 
$\rho _{\Lambda}/ \rho _{\mathrm{M}} \propto a^{3}$, that 
observers {\it today} should just happen to live right around 
the time (within a factor of $\sim$$2$ in scale factor) when 
$\Lambda$ began dominating the cosmic evolution. 

One possible way of ameliorating such problems is to give the 
Dark Energy an evolving equation of state, $w(z)$, where such 
coincidences are eliminated or moderated (like in tracker quintessence; 
e.g., \citet{ZlatevQuint}). But moving away from a Cosmological Constant 
by giving the Dark Energy an evolution in $w(z)$ also opens the door 
to other possible dynamics -- such as the ability for the Dark Energy 
to be mobile and thus perhaps to become spatially inhomogeneous 
\citep{CaldDDEsNotSmooth}. A dynamical Dark Energy (DDE) of this 
type is liable to join in structure formation to an observationally 
unacceptable degree, thus ruining it as a candidate to serve as the 
``smoothly-distributed'' missing cosmic ingredient. 

From the $1^{\mathrm{st}}$ Law of Thermodynamics, $\mathrm{Pressure} 
= - \partial [\mathrm{Energy}] / \partial [\mathrm{Volume}]$, 
so that a `negative pressure' ($P < 0$) substance like Dark Energy 
will have $\partial E / \partial V > 0$; consequently, it should be 
{\it self-attractive} on local scales (and {\it relativistically} so, 
since $w \equiv P/ \rho \sim -1$), in stark contrast to how Dark Energy 
is popularly described as a ``repulsive" or ``antigravity" material 
\citep[e.g.,][]{HeavensRepuls}. This potential instability of DDE to 
spatial perturbations must be countered 
through the addition of some nonadiabatic support pressure 
(an adiabatic positive pressure term would cause cosmic 
{\it deceleration}); and though this can apparently be done 
\citep[e.g.,][]{CaldDDEsNotSmooth,HuDEint05}, other results exist 
\citep{MaorLahavVirial} indicating that a Dark Energy 
allowed to dynamically participate in virialization does indeed show 
signs of providing an additional attractive force for clustering, 
rather than opposing clustering or leaving it unaffected. In any case, 
an inhomogeneous DDE supported against extensive clumping by 
nonadiabatic pressure would not be a perfect fluid, and since a perfect 
fluid approximation is the basis of the argument for acceleration 
($\ddot{a} > 0$) from $P < 0$ \citep[e.g.,][pp. 48-50]{KolbTurner}, 
the entire situation is best avoided if possible. 

Interestingly, the self-attractive nature of negative pressure presents 
us with the seed of an alternative idea: in its own way, {\it normal 
gravitational attraction} represents a form of `negative pressure' -- 
in particular, the gravitational attraction responsible for the growth 
of clustering in large-scale structure formation. Making a virtue of 
necessity, one may therefore recruit cosmic structure formation as the 
driver (not just the trigger, as in tracker quintessence) of the 
cosmic acceleration. This would completely solve the Coincidence Problem, 
since the remarkably contemporary onsets of both the acceleration and the 
existence of galaxies (and hence planets, life, and astronomical observers) 
would be nothing more coincidental than finding two neighboring apples 
that have fallen from the same tree. 

Such considerations are what initially led this author to search for a 
structure-formation-induced solution to the cosmic acceleration problem 
\citep{Bochner21Texas}, though a plausible acceleration mechanism was 
not yet apparent. During the past few years, a number of other researchers 
have similarly proposed a variety of methods \citep[e.g.,][as a few 
examples]{SchwarzFriedFail,RasanenFriedFail,KMNRinhomogExp,WiltInhomogNoDE} 
also attempting to obtain the observed cosmic acceleration from the breaking 
of FLRW homogeneity and the formation of structure, without requiring 
the intervention of non-standard gravitational or particle physics, 
or the non-Copernican approach of placing us near the center of a 
large cosmic void. This type of approach has become generally known as 
``backreaction". The backreaction paradigm, in the consensus view, has 
been unable to substantially replace Dark Energy as the source of the 
apparent acceleration; a conclusion which we will argue is quite 
premature. 

In this paper, in fact, we will present a physically plausible (though 
phenomenological) mechanism that is not only able to reproduce the 
cosmic acceleration as indicated by observations of standard candles 
such as Type Ia supernovae, but which can also reproduce several 
other key features of the conventional $\Lambda$CDM concordance. 

The paper will be organized as follows: in 
Section~\ref{SecBackReactImpact}, we explain why several oft-quoted 
limitations of the backreaction mechanism are inapplicable in realistic 
situations, and then use such arguments as guideposts towards developing 
our formalism in Section~\ref{SecFormalismAbbrev}; in Section~\ref{SecResults}, 
we describe our specific models and present our simulated Hubble 
curves; in Section~\ref{SecNewConcordObsTests}, we evaluate the success 
of our models in formulating an alternative ``Cosmic Concordance"; and we 
conclude in Section~\ref{SecConcludeAndSummary}. Note that the in-depth 
technical details of the research introduced here are discussed much more 
fully in a companion paper by this author \citep{BochnerAccelPaperI}.

\section{\label{SecBackReactImpact}THE COSMOLOGICAL IMPACT 
OF ``BACKREACTION"} 

At present, backreaction is not widely considered to 
be a viable method for achieving the observed acceleration 
\citep[e.g.,][]{SchwarzBackReactNotYet}. There are a number of 
important reasons for this view, reasons which seem on the surface 
to be strong arguments; but we will explain here why despite being 
technically correct, several of those arguments do not properly 
apply to a physically realistic cosmological model. 

One popular approach for estimating (and placing limits upon) the 
effects of backreaction is to use ``Swiss-Cheese'' solutions -- regular 
FLRW cosmologies with spherical `holes' cut out of them, with precisely 
designed boundary conditions so that the exterior universe (the 
`cheese') remains completely unaffected by the holes. Such models are 
advantageous because exact solutions may be used for the clumped material 
inside the holes, allowing for a precise analysis. But the weakness of 
such models is their significant lack of realism: not only do most of 
them lack any mechanism for representing structure virialization and 
stabilization, but they have in fact been specially designed to set the 
actual backreaction (as opposed to purely observational effects) to zero, 
by hermetically sealing off each condensed structure so that it cannot 
affect the exterior universe (even shielding the holes from one another). 
In the real universe, though, the `spheres of influence' of individual 
mass overdensities continually expand over time, affecting more and more 
distant objects, until multiple `independent' influences eventually 
{\it overlap and merge}. This superposition of gravitational perturbations 
from many different inhomogeneities is a phenomenon that Swiss-Cheese 
models cannot represent at all. Thus we understand why Swiss-Cheese 
backreaction calculations fail to find effects strong enough to produce 
the observed acceleration -- such as, for example, the findings of 
\citet{BisNotLTBneglig} showing that the perturbative effects of a hole 
on observations are strongly suppressed, by a factor $(L/R_{H})^{3}$ for 
passage through the hole, where $L$ is the size of the hole and $R_{H}$ is 
the Hubble radius. Recognizing the improperly artificial act of abruptly 
cutting off the gravitational influence of an overdensity at hole `edge' 
$r = L$, as the Swiss-Cheese models do, we see therefore that the smallness 
of the suppressive term $(L/R_{H})^{3}$ is itself completely artificial. 
A realistic backreaction model would not have any predetermined limits 
on $L$, but rather would allow it to grow without bound (within causal 
speed limits for the propagation of gravitational perturbation information), 
overlapping with more and more distant holes, until the factor 
$(L/R_{H})^{3}$ is no longer much of a `suppression', after all. 

Another argument claiming to limit the efficacy of backreaction is the 
acceleration ``no-go" theorem of \citet{HirataSeljak}, based upon the 
Raychaudhuri equation \citep{HawkingEllis}, demonstrating that a `true' 
(volumetric) acceleration of a spatial region is impossible in the absence 
of a material with sufficiently negative pressure ($w < -1/3$) to serve as 
Dark Energy. This conclusion only holds, however, as long as one assumes 
the cosmic matter to be irrotational -- that is, possessing zero vorticity. 
But here we must ask: {\it Why is the vorticity set to zero?} This seems 
to be an especially unusual prescription in a universe where nearly 
everything is rotating or revolving about something. Essentially all 
virialized structures in the universe larger than individual solid objects 
are stabilized against collapse by {\it some} version of vorticity or 
velocity dispersion; and it is certainly a bad physical approximation to 
take the dominant physical force (in opposition to gravity) which regulates 
cosmic structure formation, and completely set it to zero. Since vorticity 
in significant amounts -- on large scales, at least -- is not a prediction 
of inflation \citep{HirataSeljak}, it is typically regarded as a 
``small-scale player'' \citep[e.g.,][]{BuchertDEstructStatus}, relevant 
only for cosmic averages performed over domains that are on or below 
the scales of galaxy clusters; and thus it is usually set to zero for 
technical convenience \citep[e.g.,][]{RasanenZeroVort}, a seemingly 
harmless simplification. But this ignores the fact that the term actually 
appearing in the Raychaudhuri equation is not the vorticity tensor 
$\omega_{\mu \nu}$, but the {\it square} of the vorticity, $\omega^{2}$, 
a quantity that will {\it never} average away. As \citet{BuchertEhlers97} 
put it, cosmic averages of positive semi-definite quantities like 
$\langle \omega^{2} \rangle$ get ``frozen" at the size of 
``typical subdomains"; and thus they cannot be made to go to zero 
by averaging over larger domains, even if the spatially-averaged value 
of the parent tensor, $\langle \omega_{\mu \nu} \rangle$, does itself 
become negligible on large scales. Thus dropping vorticity entirely 
from the cosmic evolution is a very poor approximation. 

We do not attempt here to explicitly model the enormously complex 
backreaction effects of vorticity during the dynamical virialization 
of a structure; rather, we exploit the very simple natures of both the 
beginning state (nearly perfectly smooth FRW matter), and the ending state 
(reasonably randomly distributed Newtonian mass concentrations) of structure 
formation. This will lead (in Section~\ref{SecFormalismAbbrev}, below) to a 
phenomenological formalism depending only upon the general evolution rate 
of the cosmic matter from ``smoothness'' to ``clumpiness'' over time. 

Lastly, note that our formalism uses the backreaction effects of 
{\it Newtonian-strength} (i.e., gravitationally linear) perturbations, 
despite the extensive work of Buchert and collaborators 
\citep[e.g.,][]{BuchertEhlers97} seemingly demonstrating that the 
entire Newtonian-level backreaction can be expressed mathematically 
as a total divergence, thus guaranteeing it to be either extremely 
small or identically zero (depending upon the boundary conditions). 
But comparing Equations 1a--d of \citet{BuchertEhlers97} to 
Maxwell's Equations \citep[e.g.,][]{JacksonEM} shows that their 
formalism contains a crucial simplification: a dropping of all 
``gravitomagnetic" fields from their calculations. Similar to how 
structure formation calculations are based upon the Poisson equation, 
$\nabla^2 \Phi = 4 \pi G \rho$ (though suitably modified for the cosmic 
expansion), in which the time-derivative terms of the full wave equation 
have been dropped, this is an explicitly non-causal formalism. 
(Other backreaction no-go arguments like that in \citet{IshWald} 
are also reliant upon dropping terms like 
$\vert \partial \Phi / \partial t \vert ^{2}$, 
as seen via their Equation~2b.) 
This simplification is typically justified by citing the smallness 
of $v^{2}/c^{2}$ given the (reasonable) assumption of nonrelativistic 
speeds for most matter flows; but there is a crucial difference between 
``small" and ``suppressed". Many individually small contributions 
can be added together to produce a large overall result, if one has 
enough of them; and though Newtonian-level perturbations weaken as 
$\sim$$1/r$, the number of them in a spherical shell increases 
as $\sim$$r^{2}$ (easily overwhelming any factors of $v^{2}/c^{2}$), 
creating a total perturbative effect that would formally 
be {\it infinite} when integrated out to $r = \infty$, 
if not reined in by the finite causal horizon out to which an 
observer can `see' clumped structure that has had sufficient 
time since the Big Bang to form; a situation reminiscent 
of Olbers' Paradox \citep{WeinbergGravCosmo}.

For a causality-respecting approach, one must instead 
(as in electrodynamics) use the full wave equation, for 
special-relativistically consistent 
perturbation potential function $\Phi _{\mathrm{SR}}$: 
\begin{equation}
\nabla^2 \Phi _{\mathrm{SR}} - 
\frac{1}{c^2} \frac{\partial^2 \Phi_{\mathrm{SR}}}{\partial t^2} 
= 4 \pi G \rho ~ . 
\label{EqnPoissonDynamic}
\end{equation}
(Note that factors relating to the cosmic expansion are still 
neglected here, for simplicity.) Now, the usual impulse is to 
immediately drop the extra term in Equation~\ref{EqnPoissonDynamic}, 
involving $\partial^2 \Phi_{\mathrm{SR}} / \partial t^2$ -- equivalent 
to dropping the gravitomagnetic terms, as is done in the Buchert 
formalism -- because of its resulting prefactor of $v^{2}/c^{2}$; 
this factor would seem to make it very small given the assumption 
of nonrelativistic speeds for most matter flows, and thus (assumedly) 
ensuring it to be negligible compared to the spatial variations 
term in any backreaction calculation. But this thinking is based 
only upon considerations of individual Fourier perturbation modes, 
not on the overall causal behavior of information flow in the 
structure-forming universe. If we instead {\it keep} all terms, 
and solve Equation~\ref{EqnPoissonDynamic} as-is, then one gets 
\citep[adapting from][eq. 6.69]{JacksonEM}: 
\begin{equation}
\Phi_{\mathrm{SR}}({\bf x},t) = 
-G \int^{\infty} \frac{[\rho({\bf x^{\prime}},t)]_{\mathrm{ret}}}
{\vert {\bf x} - {\bf x}^{\prime} \vert} d^{3}x^{\prime} ~ , 
\label{EqnSRpotential}
\end{equation}
where the bracketed numerator is always evaluated at the 
{\it retarded time}, $t^{\prime} = 
t - \vert {\bf x} - {\bf x}^{\prime} \vert / c$.
It is this retarded-time condition which restores causality, 
allowing different regions of the universe to communicate 
with (and gravitationally perturb) one another; and which  
provides the escape route from the Buchert suppression of 
Newtonian-level backreaction, because such backreaction is 
{\it not} truly expressible as a total divergence. We will 
refer to this propagation of gravitational perturbation 
information between distant (though communicating) regions 
as ``causal updating".

Given the fact that the key metric perturbation function, 
$\Phi _{\mathrm{SR}} (t)$, is predominantly affected by 
inhomogeneity information coming in from distant locations, 
the retarded-time condition of an integrated formula like 
Equation~\ref{EqnSRpotential} (suitably modified for 
cosmological calculations) implicitly gives it the ability 
to exhibit relativistic behavior in what would otherwise appear 
to be nonrelativistic situations. Consider the farthest 
distance (at some given observation time, $t$) out to which 
the look-back time still represents a fairly clumpy universe; 
this distance (as a function of $t$) represents the edge of an 
outgoing ``wave of observed clumpiness" expanding outward from 
an observer at $c$, which ensures that the potential function 
$\Phi _{\mathrm{SR}}$ for perturbations felt by that observer 
will be fully (special-)relativistic in character, {\it even if 
none of the matter in the universe is actually moving at 
relativistic speeds}\footnote{As an analogy, consider the 
contact point between the two blades of a very long pair of 
scissors. The rate at which this contact point moves outward 
from the central pivot does not represent the physical motion 
of any real object, and hence is not limited by the speed 
of the material in the blades as they come together.}. This 
unexpected result is due to the fact that cosmological systems 
(unlike anything else) are infinite in size, meaning that the 
inherently relativistic act of causal observation will end up 
encompassing spatial volumes which are incomparably vast, 
causing even small effects to sum together to produce a 
dominant overall influence. Perhaps surprisingly, therefore, 
the apparent acceleration of the universe which we 
observe is not caused by any individually powerful or 
gravitationally exceptional specific objects or regions, but 
rather is an effect that comes from the summed influences of the 
weak-gravity Newtonian tails of innumerable mass concentrations, 
imposing their combined effects upon us from extraordinarily 
large distances.

In this approach, it is clear that we are advocating a 
diametrically-opposed view from that of other authors who have 
sought to define a ``finite infinity'' 
\citep[e.g.,][]{DPGCoxHowFarInfinity}, representing a very limited 
boundary from within which significant effects upon some specified 
local volume can have ever arrived. Such a limitation would restrict 
the history of `important' interactions to be within an effective 
``matter horizon'' \citep{EllisStoeBogusMatterHorizon} of only a 
few Mpc in size, delineated by the timelike world lines traveled by 
pressure-free matter due to scalar perturbations. But interestingly, 
both \citet{DPGCoxHowFarInfinity} and \citet{EllisStoeBogusMatterHorizon}
acknowledge the Great Attractor/Shapley Concentration as a likely 
source of our bulk flow -- i.e., as the completely dominant influence 
upon the motion of our entire neighborhood of galaxy clusters -- 
despite the clearly contradictory fact (for their claims) that 
those structures which control our bulk motion are exceedingly 
far beyond any reasonable estimate of our so-called matter horizon. 
Thus one cannot take at face value the assertion in 
\citet{EllisStoeBogusMatterHorizon} that, ``the strengths of any 
other possible long distance influences... gravitational waves, 
or electric Weyl tensor components from sources outside that region 
-- are insignificant compared to local effects.'' Notably, it has been 
shown by \citet{LudvigsenGWGeodesicDev} that even an arbitrarily small 
energy flux due to gravitational waves can result in a finite amount of 
``geodesic deviation'' (actually, long-term positional displacements 
of observers due to permanent metric changes) in the very distant 
radiation zone. 

Nevertheless, there are very straightforward calculations using 
gravitational perturbation theory which are commonly believed 
to contradict any assertions of the significance of such 
nonlocally-acting causal backreaction effects. The problem, however, 
is the fact that perturbation theory is singularly ill-equipped for 
dealing with the most important perturbative effects in an unbounded 
system like the (effectively) infinite universe. The 
central difficulty with perturbation theory here relates to its 
basic program of identifying the `important' physics by labeling 
the amplitude of each term as ``large'' or ``small'', and then 
dropping the small terms in order to focus upon the large ones. 
For example, \citet{KMNRinhomogExp} neglect information-carrying 
tensor modes generated by nonlinear scalar perturbations 
(i.e., virializing masses) in the final expressions for their analysis 
(as well as assuming irrotational dust, thus neglecting vorticity). 
In a similar vein, both \citet{RasanenPertPropModes} and 
\citet{GromesSmallBackreaction} assume that time derivatives 
of metric perturbations are small, leading to the conclusion 
that velocity-dependent terms can safely be neglected; and 
\citet{GromesSmallBackreaction} specifically uses this point as a 
primary argument for dismissing the effects of causal backreaction 
entirely. Obviously, though, if one removes all ``causal'' aspects 
from ``causal backreaction'', then nothing will be left, and it is 
unsurprising for one to then find that backreaction fails to work, 
given that the principal physics responsible for it has been 
intentionally set to zero. But this basic procedure is conceptually 
flawed in the case of causal backreaction, which deals with 
propagating modes, since such modes can bring in gravitational 
perturbation information to a local observer from vast spatial volumes, 
so that a term with a `small amplitude' can actually produce an 
enormous overall effect. 

The habit of conflating ``small'' with ``negligible'' is a 
continuing error in perturbation theory analyses of cosmological 
evolution, since the {\it size} of these effects, in terms of amplitude, 
effectively does not matter: it is practically irrelevant how small the 
time-derivative terms may be, when one realizes that such effects are 
being cumulatively contributed by {\it all} of the matter within a 
causal horizon that may be billions of light-years in extent, thus 
multiplying that amplitude by an amount of mass easily large enough 
to overcome its inherent smallness. Furthermore, even if the amplitudes 
of the relevant perturbative terms for causal backreaction were somehow 
`magically' made even smaller than they naturally are, this still would 
not shut off the causal backreaction effect, but merely postpone it to 
begin a little bit later, since a somewhat larger causal horizon of 
self-stabilizing inhomogeneities would then be needed to supply a 
large enough integrated effect to cause an apparent acceleration. 
In that sense, one could {\it never} make the amplitudes of those 
perturbation terms small enough to avoid the eventual dominance of 
causal backreaction, since a big enough causal horizon (containing 
a sufficiently large amount of virializing mass) can always be 
attained after a sufficiently long period of time, which when 
multiplied by even the smallest amplitude, would eventually produce 
a term of order unity that then proceeds to dominate the cosmic 
evolution. Thus for causal backreaction in an effectively infinite 
universe, {\it there is no such thing as ``negligible''.}

\section{\label{SecFormalismAbbrev}THE PHENOMENOLOGICAL 
FORMALISM OF CAUSAL BACKREACTION}

We require a practical way to implement these ideas, 
so consider the following. It is well known 
\citep[][pp. 474-475]{WeinbergGravCosmo} that the Friedmann 
expansion for some spherical volume $\mathbf{V}$ can be 
derived -- using nonrelativistic Newtonian equations, in fact 
-- without reference to anything outside of that sphere. In 
a sense, the outside universe might as well not even exist for 
$\mathbf{V}$, in a completely smooth cosmology. 

This changes when perturbations develop, however, since clumped 
structures external to $\mathbf{V}$ exert measurable gravitational 
forces upon the mass within it; and such forces are {\it new}, as if 
those clumped structures were brought in from infinity at some recent 
time during the structure formation epoch, to only then begin affecting 
$\mathbf{V}$. But one cannot feel a gravitational {\it force}, unless 
one simultaneously feels a gravitational {\it potential}; and so the 
perturbing potential within $\mathbf{V}$ must also be new in every 
physically measurable sense as well, approaching and entering 
$\mathbf{V}$ in causal fashion from some clumped object (and from 
all others) as they develop over time, everywhere in the 
surrounding universe. 

For a randomly-distributed collection of inhomogeneities, the forces 
exerted upon the mass in $\mathbf{V}$ would pull in all directions, 
largely canceling one another; and indeed, our formalism is based 
upon a ``smoothly-inhomogeneous'' universe for which spatial variations 
are essentially averaged out. But what does {\it not} cancel out, 
regardless of how the perturbations outside of $\mathbf{V}$ are 
distributed -- because it is a non-directional scalar, rather than a 
vector -- is their combined contribution to the absolute level of 
the potential $\Phi$, which may not matter in true Newtonian physics 
(only potential differences or gradients do), but which does matter 
in general relativity. This overall perturbation potential 
$\Phi _{\mathrm{SR}} (t)$ is treated here as independent of position; 
but it fully obeys causality, growing as perturbation information 
comes in from the limits of one's observational horizon. Our 
phenomenological approach thus models the inhomogeneity-perturbed 
evolution of (any volume) $\mathbf{V}$ with a metric that contains 
the individual Newtonian-level perturbations from all clumped, 
virialized structures that have been causally `seen' within 
$\mathbf{V}$ by time $t$, superposed {\it on top of} its 
(internally-generated) background Friedmann expansion. 

This is clearly a very heuristic picture, with the growing perturbation 
potential $\Phi _{\mathrm{SR}} (t)$ not representing the {\it literal} 
metric of the universe, but rather acting as a quantifiable shorthand 
for the dominant backreaction effects from structure formation. 
What is really happening, is that virializing overdensities halt 
their collapse by concentrating their local vorticity (and/or 
velocity dispersion); this concentrated vorticity leads to real, 
extra volume expansion, representable (in the final state) at 
great distances by the tail of a Newtonian perturbation potential 
to the background FRW metric; this Newtonian tail propagates outward 
into space by inducing inward mass flows towards the virialized object 
from farther and farther distances as time passes; and the total 
perturbation at any given location will be the combined effects of 
innumerable Newtonian tails of this type, coming in causally 
(from all directions) from within that location's ever-expanding 
``inhomogeneity horizon''. Our formalism merely uses 
$\Phi _{\mathrm{SR}} (t)$ to represent all of these complex 
physical processes in a very simple way, and we will see 
that this is sufficient for producing an observed acceleration 
fully in agreement with observations. 

To phenomenologically model the cosmological phase transition from 
the extremely symmetrical and smooth (``unclumped") state during the 
early universe, to its almost entirely ``clumped" state today, 
we employ a ``clumping evolution function", $\Psi (t)$, defined 
over the range from $\Psi (t) = 0$ (perfectly smooth matter) to 
$\Psi (t) = 1$ (everything clumped). Note that we assume a spatially 
flat background -- for the pre-perturbed universe -- everywhere in 
our calculations, and that we therefore assume the entire universe to 
be composed of pressureless dust with $\Omega^\mathrm{FRW} _\mathrm{M} 
\equiv 1 \ne \Omega^\mathrm{Obs} _\mathrm{M}$ (the superscript ``Obs'' 
referring to observational quantities). 

The Newtonian approximation for a single clumped object of mass $M$ 
(representing a virialized, self-stabilized inhomogeneity), 
embedded at the origin ($r = 0$) in an expanding, spatially flat, 
matter-dominated (MD) universe can be obtained by linearizing the 
McVittie solution \citep{McVittieBHinFRW}, as can be seen from the 
perturbed FRW expression given in \citet{KaloKlebanBHinFRW}. One gets: 
\begin{equation}
ds^2 \approx - c^{2} [1 + (2/c^{2}) \Phi (t)] d t^{2} 
+ [a_{\mathrm{MD}}(t)]^{2} [1 - (2/c^{2}) \Phi (t)] d r^{2}
+ [a_{\mathrm{MD}}(t)]^{2} r^{2} [d {\theta}^{2} 
                           + \sin^{2}{\theta} d {\phi}^{2}] 
~ , 
\label{NewtPertSingleClump}
\end{equation} 
where $\Phi (t) \equiv \{ - G M / [a_{\mathrm{MD}}(t) r] \}$, 
and $a_{\mathrm{MD}}(t) \propto t^{2/3}$ is the 
unperturbed MD scale factor. 

Assuming that such mass concentrations will be randomly and 
(sufficiently) evenly distributed throughout the universe, 
one must integrate over distance (from a particular 
observer's location) for the various clumps, and one must 
also {\it angle-average} over them to get the metric for 
the combined effects that would represent an averaged $ds^2$ 
typical for that observer experiencing displacements in 
any given direction. Since a displacement of magnitude 
$\vert d \vec{r} \vert$ will have a radial component 
(with respect to any particular mass concentration) of 
$\vert d \vec{r} \vert \cos \theta$, angle-averaging thus 
brings in a factor of $\langle \cos^{2} \theta \rangle = (1/3)$ 
for the spatial part of the metric perturbation term, 
while leaving its temporal part unchanged. 

Consider now the observer to be located at the origin, 
surrounded by a set of (roughly identical) discrete mass 
concentrations with {\it total} mass $M$, all located within 
a particular spherical shell at coordinate distance $r^{\prime}$ 
from the origin, but randomly distributed in $(\theta, \phi)$.
From the above arguments, we can write the total, {\it linearly} 
summed and angle-averaged metric in `isotropized' fashion, 
as: 
\begin{equation}
ds^{2} = 
- c^{2} \{ 1 - [ R_{\mathrm{Sch}}(t) / r^{\prime} ] \} ~ dt^{2} 
~ + ~ [a_{\mathrm{MD}}(t)]^{2} 
\{ 1 + (1/3) [ R_{\mathrm{Sch}}(t) / r^{\prime} ] \} 
~ \vert d \vec{r} \vert ^{2} ~ ,
\label{EqnAngAvgBHMatDomWeakFlat}
\end{equation} 
where 
$R_{\mathrm{Sch}}(t) \equiv \{ (2 G M / c^{2} ) / 
[a_{\mathrm{MD}}(t)] \}$, and 
$\vert d \vec{r} \vert ^{2} 
\equiv (d r^{2} + r^{2} d \theta ^{2} 
+ r^{2} \sin^{2}{\theta} d \phi ^{2} ) 
= \vert d \vec{x} \vert ^{2} 
\equiv (dx^{2} + dy^{2} + dz^{2})$. 
Regarding the spatial metric perturbation term as a `true' 
increase in spatial volume, and the perturbation in $g_{t t}$ 
as an `observational' slowdown of the perceptions of observers 
(relative to the expansion governed by cosmic time $t$), 
we see that even if the spatial term by itself is not enough to 
generate a real volumetric acceleration, when coupled with the 
temporal term the entire perturbation may indeed be enough to 
create a so-called ``apparent acceleration" sufficient to 
explain all of the relevant cosmological observations. 

The expression in Equation~\ref{EqnAngAvgBHMatDomWeakFlat} 
represents the perturbations to the metric due to masses at some 
specific coordinate distance $r^{\prime}$ -- and thus from a 
specific look-back time $t^{\prime}$ -- as seen from some particular 
observational point. The total metric at that spacetime point must 
be computed via an integration over all possible distances, out to 
the distance (and thus look-back time) at which the universe had 
been essentially unclustered. Finally, a light ray reaching us from 
its source (e.g., a Type Ia supernova being used as a standard candle) 
travels to us in a path composed of a collection of such points, 
where the metric at each point must be calculated via its own 
integration out to its individual ``inhomogeneity horizon''; and 
only by calculating the metric at every point in the pathway from 
the supernova to our final location here at $r = z = 0$ can we 
figure out the total distance that the light ray has been able to 
travel through the increasingly perturbed metric, given its 
emission at some specific redshift $z$. 

\begin{figure}
\begin{center}
\includegraphics[scale=0.75]{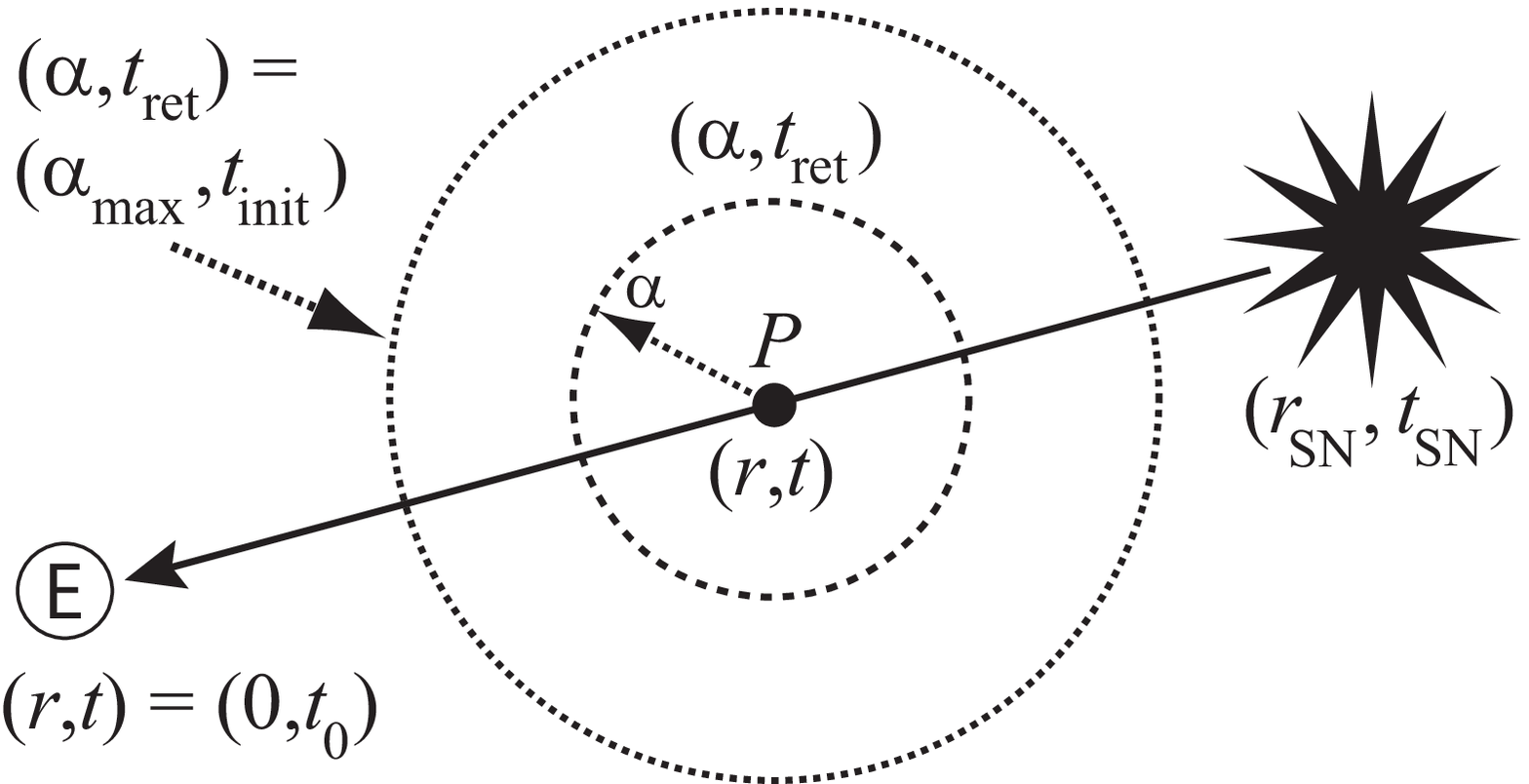}
\end{center}
\caption{Geometry for computing the inhomogeneity-perturbed 
metric at each point along the integrated path of a light ray 
from a supernova to our observation point at Earth.}
\label{FigSNRayTraceInts}
\end{figure}

Consider a light ray emitted by a supernova at cosmic 
coordinate time $t = t_{\mathrm{SN}}$, which then propagates 
from the supernova at $r = r_{\mathrm{SN}}$, to us at $r = 0$, 
$t = t_{0}$. We refer here to the geometry depicted in 
Figure~\ref{FigSNRayTraceInts}.

\placefigure{FigSNRayTraceInts}

For each point $P \equiv (r,t)$ of the trajectory, the metric 
at that point will be perturbed away from the background FRW 
form by all of the virialized clumps that have entered its 
causal horizon by that time. Consider a sphere of (coordinate) 
radius $\alpha$, centered around point $P$, with coordinates 
$(\alpha, t_{\mathrm{ret}})$ (where $t_{\mathrm{ret}} \leq t$ 
is the retarded time), defined such that the information 
about the state of the clumping of matter on that sphere 
at time $t_{\mathrm{ret}}$ will arrive -- via causal 
updating, traveling at the speed of null rays -- to 
point $P$ at the precise time $t$. To compute the 
fully-perturbed metric at $P$, we must integrate over the 
clumping effects of all such radii $\alpha$, from 
$\alpha = 0$ out to $\alpha _{\mathrm{max}}$, the farthest 
distance from $P$ out from which clumping information can 
have causally arrived since the clustering of matter had 
originally begun in cosmic history.

The full derivations of all formulae shown below are given 
in \citet{BochnerAccelPaperI}; here we quote the resulting 
equations, which are used for our causal backreaction 
simulations. 

For a FRW metric with $a(t) = a_{\mathrm{MD}}(t) 
\equiv a_{0} (t / t_{0})^{2/3} 
\equiv c [18 t^{2} / H_{0}]^{1/3}$, 
the retarded time $t_{\mathrm{ret}}$ can be given as: 
\begin{equation}
t_{\mathrm{ret}} (t, \alpha) = 
t_{0} [(t/t_{0})^{1/3} - \alpha]^{3} ~ .
\label{EqntRet}
\end{equation}
Similarly, we can determine $\alpha _{\mathrm{max}}$, 
given some initial time $t_\mathrm{init}$ at which 
structure formation can be reasonably said to have 
started (i.e., 
$\Psi(t \le t_\mathrm{init}) \equiv 0$):
\begin{equation}
\alpha _{\mathrm{max}} (t, t_\mathrm{init})
= [(t/t_{0})^{1/3} - (t_\mathrm{init}/t_{0})^{1/3}] ~ .
\label{EqnalphaMax}
\end{equation}

Now, how the metric is affected at $P$ by a spherical 
shell of material at coordinate radius $\alpha$ depends 
upon the state of clumping there at the appropriate 
retarded time: $\Psi [t_{\mathrm{ret}} (t, \alpha)]$. 
The total effect is computed by integrating all shells 
from $\alpha = 0$ out to 
$\alpha = \alpha _{\mathrm{max}} (t, t_\mathrm{init})$; 
but in order to compute the metric perturbation from 
each shell quantitatively, it is first necessary to 
relate this clumping function to an actual physical 
density of material. 

Recalling Equation~\ref{EqnAngAvgBHMatDomWeakFlat}, 
we have the perturbation term 
$[ R_{\mathrm{Sch}}(t) / r^{\prime} ] = 
\{ (2 G M / c^{2} ) / [r^{\prime} ~ a_{\mathrm{MD}}(t)] \}$, 
where the value of $M$ to use here at time $t$ is given 
by the clumped matter density, 
$[ \Psi (t) \rho _{\mathrm{crit}}(t) ]$,  
times the infinitesimal volume element of the shell. 
Collecting all relevant factors (and letting 
$t^{\prime} \equiv t_{\mathrm{ret}} (t, \alpha)$), 
the integrand at coordinate distance $\alpha$ can 
eventually be written as: 
\begin{equation}
[ R_{\mathrm{Sch}}(t) / r^{\prime} ]
_{r^{\prime} = \alpha \rightarrow (\alpha + d \alpha)} 
= \{12 ~ \Psi (t^{\prime}) 
~ [(t_{0} / t)^{2/3}]
~ {[ \alpha d \alpha} ]\} ~ , 
\label{EqnIintegrandPrelim1stLineOnly}
\end{equation}
where for simplification we have used 
$H_{0} = (2/3) t_{0}^{-1}$ and the 
fact that $[\rho(t) a(t)^{3}]$ is constant, both 
true for a matter-dominated universe (but see caveats 
below). Note also that only $\Psi$ is evaluated at the 
retarded time, $t_{\mathrm{ret}} (t, \alpha)$, since the 
only ``relativistic" piece of propagating information 
which is causally delayed is the state of clumping 
($\Psi [t_{\mathrm{ret}} (t, \alpha)]$) that has 
just then arrived from coordinate distance $\alpha$ 
to observer $P$ at $(r,t)$.

From this result, we can now determine the total 
integrated metric perturbation function due to clumping, 
$I(t)$, as experienced by a null ray (or any observer) 
passing through point $P$ at $(r,t)$: 
\begin{equation}
I(t) = 
\int^{\alpha _{\mathrm{max}} (t, t_\mathrm{init})}_{0}
\{12 ~ 
\Psi [t_{\mathrm{ret}} (t, \alpha)] 
~ [(t_{0} / t)^{2/3}] \} ~ 
\alpha ~ d \alpha ~ ,
\label{EqnItotIntegration}
\end{equation}
with $I(t)$ implicitly being a function of 
$t_\mathrm{init}$ (with $I(t) \equiv 0$ for 
$t \leq t_\mathrm{init}$), as well as of $t_{0}$.

Finally, we can insert this result back into the 
formalism of Equation~\ref{EqnAngAvgBHMatDomWeakFlat}, 
to obtain the final clumping-perturbed metric that 
we will use for all of our subsequent cosmological 
calculations: 
\begin{equation}
ds^{2} = 
- c^{2} [ 1 - I(t) ] ~ dt^{2} 
~ + ~ \{ [a_{\mathrm{MD}}(t)]^{2} ~ 
[ 1 + (1/3) I(t) ] \} 
~ \vert d \vec{r} \vert ^{2} ~ . 
\label{EqnFinalBHpertMetric}
\end{equation}

Note that we have implicitly made a significant approximation 
in this above derivation, in that several quantities (e.g., 
$t_{\mathrm{ret}}$, $\alpha _{\mathrm{max}}, 
[ R_{\mathrm{Sch}}(t) / r^{\prime} ]$) were calculated 
relative to the {\it unperturbed} background metric, rather than 
with respect to the inhomogeneity-perturbed metric itself. 
This key approximation is the dropping of what we refer to as 
``recursive nonlinearities'' -- called that because the integrated 
function $I(t)$ invariably depends upon itself recursively, in an 
operationally nonlinear way -- and this simplification is done here 
out of practical necessity for this initial, proof-of-principle work; 
but a fully self-consistent treatment would need to account for 
how causal updating is itself slowed by the metric perturbation 
information carried by causal updating, since the strength of the 
acceleration as $z \rightarrow 0$ should be moderated to some extent 
by these recursive nonlinearities. Incorporating them properly 
represents a crucial next step for our causal backreaction formalism; 
but for now, we simply note that the output cosmological fits and 
parameters from our calculations here will unavoidably possess some 
systematic theoretical uncertainty. 

With Equation~\ref{EqnFinalBHpertMetric} as our metric, 
we must relate unperturbed (``FRW'') quantities -- such as 
$z^\mathrm{FRW}(t) \equiv [ (t_{0}/t)^{2/3} - 1 ]$ -- to 
actual physically-observed quantities, such as the corresponding 
observational redshift, given as follows: 
\begin{equation}
z^\mathrm{Obs}(t) \equiv 
\frac{\sqrt{g_{r r}(t_{0})}}{\sqrt{g_{r r}(t)}} - 1 
= [ \sqrt{\frac{1 + (1/3) I(t_{0})}{1 + (1/3) I(t)}} 
~ (t_{0}/t)^{2/3}] - 1 ~ .
\label{EqnDefNzObs}
\end{equation}

In order to compute the supernova-based luminosity distance 
function (i.e., the observable curve which most directly traces 
out the cosmic expansion history), we require a formula for the 
coordinate distance $r$ of a supernova going off at coordinate 
time $t$, which would be seen on Earth precisely now; we find 
that the resulting expression is: 
\begin{equation}
r^\mathrm{FRW}_{\mathrm{SN}}(t) = 
\frac{c}{a_{0}} \int^{t_{0}}_{t}
\{ (t_{0}/{t^{\prime}})^{2/3} 
~ \sqrt{\frac{1 - I(t^{\prime})}{1 + 
(1/3) I(t^{\prime})}} \} 
~ d t^{\prime} ~ . 
\label{EqnRofTresult}
\end{equation}
This formula can then be converted into an expression for the 
observed luminosity distance function, ultimately yielding: 
\begin{equation}
d_{\mathrm{L}, \mathrm{Pert}}(t) = 
\frac{1 + (I_{0}/3)}{\sqrt{1 + [I(t_{r}) / 3]}}
~ 
\frac{c ~ t_{0}}{t_{r}^{2/3}} 
~ 
\int^{1}_{t_{r}}
\{ (t^{\prime}_{r})^{-2/3} 
~ \sqrt{\frac{1 - I(t^{\prime}_{r})}{1 + 
[I(t^{\prime}_{r}) / 3]}} \} 
~ d t^{\prime}_{r} 
~ , 
\label{EqnDlumDefnResult}
\end{equation} 
where $I_{0} \equiv I(t_{0})$, and $t_{r}$, 
$t^{\prime}_{r}$ are dimensionless time ratios 
(e.g., $t_{r} \equiv t / t_{0}$), with no change 
to the essential form of $I(t)$ (i.e., 
$I(t) = I(t_{r} \cdot t_{0}) \Rightarrow I(t_{r})$). 
Finally, this luminosity distance function 
$d_{\mathrm{L}, \mathrm{Pert}}$ computed for any 
clumping evolution model $\Psi (t)$ is converted to the 
form $\Delta (m - M)_{\mathrm{Pert}} (z^\mathrm{Obs})$ 
by subtracting off the $d_{\mathrm{L}}$ function for 
an empty, coasting universe; and the resulting curve 
can be plotted as a residual Hubble diagram, for use 
in comparisons against theoretical FLRW models and 
standard candle data from supernova observations.

\section{\label{SecResults}MODELS, PARAMETERS, AND RESULTS}

We must design a set of clumping evolution functions, 
$\Psi (t^\mathrm{FRW})$, to model the overall combined effect 
due to the linear evolution, the nonlinear regime, and the final 
virialization of individual clustering masses, along with the 
triggering (often via collisions) of newly-forming clumps. 

Observational data and simulations are not sufficiently constraining, 
so we opt here for simplicity. Considering different varieties of 
time-dependence, we choose: $\Psi(t) \propto t^{2/3}$, proportional 
to the linear density contrast evolution in a matter-dominated universe 
\citep[e.g.,][]{KolbTurner}; and $\Psi(t) \propto t$, the amount 
of clumping being simply proportional to the time that has elapsed. 
We also try an `accelerating' clumping rate, $\Psi(t) \propto t^{2}$, 
potentially corresponding to the final {\it nonlinear} evolution of a 
density perturbation \citep[][p. 322]{KolbTurner}; though the lesser 
amount of clumping at early times for this model will lead to 
comparatively weak causal backreaction effects. Thus we define: 
\begin{mathletters}
\begin{eqnarray}
\Psi _{\mathrm{MD}} (t) & \equiv & 
\Psi _{0} ~ 
(\frac{t - t_\mathrm{init}}{t_{0} - t_\mathrm{init}})^{2/3}
\\
\Psi _{\mathrm{Lin}} (t) & \equiv & 
\Psi _{0} ~ 
(\frac{t - t_\mathrm{init}}{t_{0} - t_\mathrm{init}}) 
\\
\Psi _{\mathrm{Sqr}} (t) & \equiv & 
\Psi _{0} ~ 
(\frac{t - t_\mathrm{init}}{t_{0} - t_\mathrm{init}})^{2}
~ , 
\end{eqnarray}
\label{EqnClumpModels}
\end{mathletters}
where $t_\mathrm{init}$ represents the effective beginning 
of clumping (i.e., $\Psi(t \le t_\mathrm{init}) \equiv 0$), and 
$\Psi _{0}$ represents the ultimate degree of clumping today. 

Appropriate values for the two adjustable parameters, 
$z_\mathrm{init}$ ($= [ (t_{0}/t_\mathrm{init})^{2/3} - 1 ] $) 
and $\Psi _{0}$, are chosen via astrophysical considerations. 
First, we link the beginning of structure formation to 
cosmological reionization. Since \citet{WMAP5yrLikeliParams} 
supports a possibly extended period of partial reionization 
lasting from $z \sim 20 - 6$, we conservatively bracket this by 
using $z_\mathrm{init} = (5,10,15,25)$ for our simulation runs. 

To specify $\Psi _{0}$, we take typical concordance 
values such as $\Omega^{\mathrm{Obs}}_{\mathrm{M}} 
\equiv 1 - \Omega^{\mathrm{Obs}}_{\Lambda} \sim 0.27$, 
with $\Omega^{\mathrm{Obs}}_{b} \sim 0.04$ 
(thus $\Omega^{\mathrm{Obs}}_{\mathrm{DM}} \sim 0.23$), 
and scale them up to flatness without Dark Energy. This yields 
$\Omega^{\mathrm{FRW}}_{b} \sim [ 0.04 (1.0/0.27) ] 
\simeq 0.15$, and $\Omega^{\mathrm{FRW}}_{\mathrm{DM}} 
\simeq 1 - \Omega^{\mathrm{FRW}}_{b} \simeq 0.85$. Estimating the 
various fractions of dark and baryonic matter which may currently 
be clumped versus unclumped \citep{BochnerAccelPaperI}, we choose 
the range $\Psi _{0} = (0.78,0.85,0.92,0.96,1.0)$ for our runs. 

\begin{figure}
\begin{center}
\includegraphics[scale=1.0]{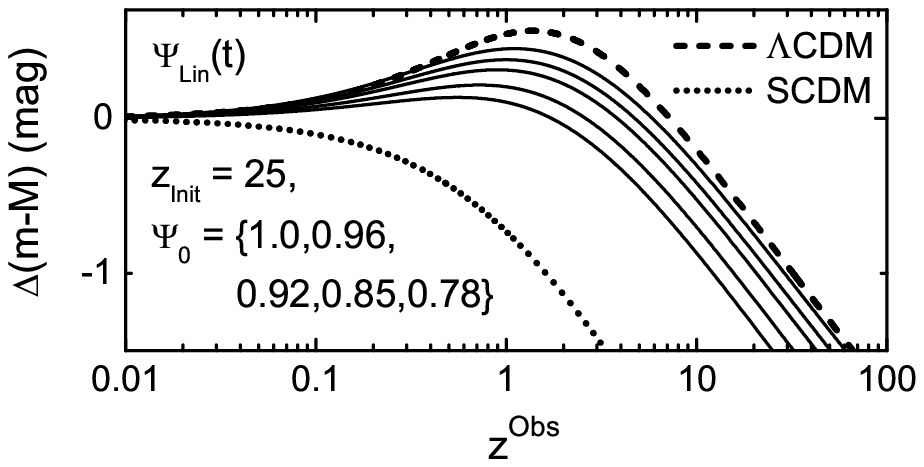}
\end{center}
\caption{Residual Hubble diagrams for selected 
$\Psi _{\mathrm{Lin}} (t)$ clumping functions 
(solid lines), plotted versus flat $\Lambda{\mathrm{CDM}}$ 
($\Omega _{\Lambda} = 0.73 = 1 - \Omega _\mathrm{M}$) 
and SCDM ($\Omega_{\mathrm{M}} = 1$) cosmologies.}
\label{FigLinPlot}
\end{figure}

\begin{figure}
\begin{center}
\includegraphics[scale=1.0]{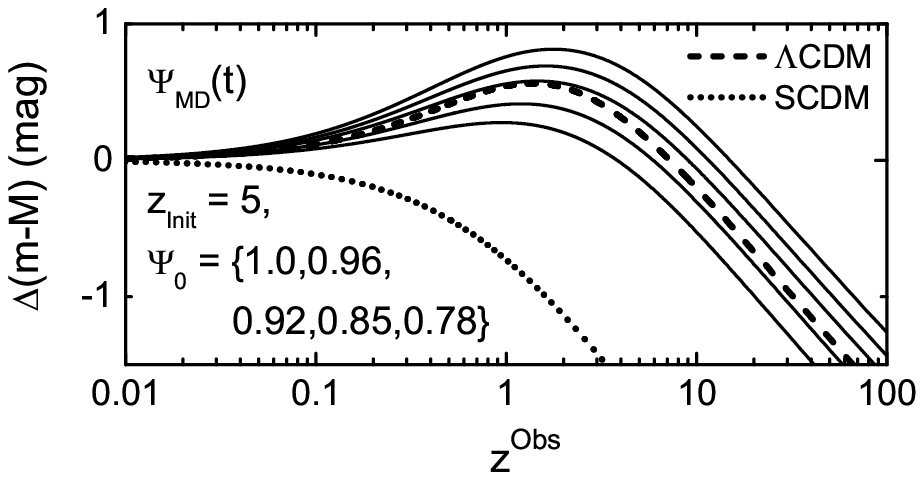}
\end{center}
\caption{Residual Hubble diagrams for selected 
$\Psi _{\mathrm{MD}} (t)$ clumping functions 
(solid lines), plotted versus flat $\Lambda{\mathrm{CDM}}$ 
($\Omega _{\Lambda} = 0.73 = 1 - \Omega _\mathrm{M}$) 
and SCDM ($\Omega_{\mathrm{M}} = 1$) cosmologies.}
\label{FigMDPlot}
\end{figure}

Four different $z_\mathrm{init}$ values, five different $\Psi _{0}$ 
values, and three different clumping evolution models gives us 
$4 \times 5 \times 3 = 60$ simulation runs in total. Residual Hubble 
diagrams of these model cosmologies show them to be broadly successful 
in reproducing a Cosmological-Constant-like observed acceleration, 
with the $\Psi _{\mathrm{Sqr}}$ runs resembling flat $\Lambda$CDM 
with $\Omega_{\Lambda} \sim 0.3-0.4$, the $\Psi _{\mathrm{Lin}}$ runs 
resembling $\Omega_{\Lambda} \sim 0.5-0.8$, and the $\Psi _{\mathrm{MD}}$ 
runs resembling $\Omega_{\Lambda} \sim 0.65-0.97$. Qualitatively, 
a $\sim$dozen of these model cosmologies fit the data essentially as 
well as the best-fit flat $\Lambda$CDM model. Two panels containing 
several of these best runs are shown in Figures~\ref{FigLinPlot} 
and \ref{FigMDPlot}.

\placefigure{FigLinPlot}

\placefigure{FigMDPlot}

For our quantitative analysis below, we fit the 
residual distance modulus function for each model, 
$\Delta \mu_{\mathrm{Pert}} \equiv \Delta (m - M)_{\mathrm{Pert}}$, 
to the SCP Union compilation \citep{KowalRubinSCPunion} 
of 307 Type Ia supernovae (SNe). Using their publicly available 
$\mu _{\mathrm{SN}, i}$ and $\sigma _{\mathrm{SN}, i}$ data, 
we compute values of $\chi ^{2} _{\mathrm{Fit}}$ for these 
$\Delta \mu_{\mathrm{Pert}}$ curves, minimized for each run 
by optimizing $H^{\mathrm{Obs}}_{0}$. Given these results, 
the corresponding fit probability values ($P_{\mathrm{Fit}}$) for 
each run are then computed using the $\chi ^{2}$ distribution with 
$N_{\mathrm{DoF}} = (307 - N_{\mathrm{Fit}})$ degrees of freedom, 
where $N_{\mathrm{Fit}} = 3$ for our $\Psi (t)$ models (optimizing 
$\Psi _{0}$, $z_\mathrm{init}$, and $H^{\mathrm{Obs}}_{0}$), 
$N_{\mathrm{Fit}} = 2$ for flat $\Lambda$CDM 
(optimizing $\Omega _{\Lambda} \equiv 1 - \Omega _\mathrm{M}$ 
and $H^{\mathrm{Obs}}_{0}$), and $N_{\mathrm{Fit}} = 1$ for 
flat SCDM (optimizing $H^{\mathrm{Obs}}_{0}$ only).

\section{\label{SecNewConcordObsTests}FORGING A NEW CONCORDANCE}

Beyond just explaining the apparent acceleration seen in the SNe data, 
a true model of the universe must satisfy a variety of complementary 
constraints before it can be regarded as a fully-consistent replacement 
for the traditional $\Lambda$CDM ``Cosmic Concordance''. 

To obtain a new concordance without Dark Energy, we must relate 
the observational (``dressed'') parameters to the unperturbed FRW 
(``bare'') parameters of our models. These mathematical relationships 
are derived in \citet{BochnerAccelPaperI}; here we will simply quote 
the output results for our best-fitting models. 

The parameters studied include: $H^\mathrm{FRW}_{0}$ (versus measured 
Hubble Constant $H^\mathrm{Obs}_{0}$); the physical age of the universe, 
$t^\mathrm{Obs}_{0}$; the unperturbed cosmic density, 
$\Omega^\mathrm{FRW}_\mathrm{M}$; the observed CMB acoustic scale, 
$l^\mathrm{Obs}_{\mathrm{A}}$; the apparent (overall) cosmic equation 
of state, $w^\mathrm{Obs}_{0}$; and the observed jerk (or jolt) parameter, 
$j^\mathrm{Obs}_{0}$. Observables $H^\mathrm{Obs}_{0}$, $w^\mathrm{Obs}_{0}$, 
and $j^\mathrm{Obs}_{0}$ are computed from the Taylor series expansion 
coefficients of the luminosity distance function, 
$d_{\mathrm{L}, \mathrm{Pert}}(z^\mathrm{Obs})$, for 
$z^\mathrm{Obs} \rightarrow 0$ \citep[see, e.g.,][]{RiessGoldSilver}. 

Besides achieving an `acceleration' ($w^\mathrm{Obs}_{0} < -1/3$), our 
key result is that when perturbations exist (i.e., $I(t) > 0$), one has 
$H^\mathrm{FRW}_{0} < H^\mathrm{Obs}_{0}$, which is astrophysically 
interesting \citep[e.g.,][]{BlanchardAltConcord} and solves a number of 
problems with matter-only cosmologies. First, since 
$(t^\mathrm{FRW}_{0} , t^\mathrm{Obs}_{0}) \propto 1 / H^\mathrm{FRW}_{0}$, 
a low $H^\mathrm{FRW}_{0}$ can solve the classic Age Problem/Crisis 
in cosmology \citep[e.g.,][]{KolbTurner} by increasing $t^\mathrm{Obs}_{0}$ 
from the $\sim$$9-10$ GYr expected from always-decelerating flat SCDM 
cosmologies, to values of $\sim$$13-14$ GYr for our strongly-perturbed models. 

Secondly, since $\Omega _\mathrm{M} = \rho _\mathrm{M} / \rho _\mathrm{crit} 
\propto \rho _\mathrm{M} / H^{2}_{0}$, a universe that {\it appears} to have 
insufficient density for closure due to $\Omega^\mathrm{Obs}_\mathrm{M} 
\propto \rho _\mathrm{M} / (H^\mathrm{Obs}_{0})^{2} \sim 0.3$, may very well 
be spatially flat (in a pre-perturbed sense) for the {\it same} 
physical matter density $\rho _\mathrm{M}$, with 
$\Omega^\mathrm{FRW}_\mathrm{M} \propto \rho _\mathrm{M} / 
(H^\mathrm{FRW}_{0})^{2} \simeq 1$. (And note that CMB-related tests of 
cosmic flatness -- e.g., \citet{WMAP7yrLikeliParams} -- primarily 
measure the pre-perturbed era.) Thus we can reconcile the apparent 
contradiction between the CMB measurements of $l^\mathrm{Obs}_{\mathrm{A}}$ 
indicating flatness, and our seemingly low-matter-density universe (as 
indicated by the growth of structure), {\it without} requiring the addition 
of any non-clustering Dark Energy species to fill the apparent gap between 
$\Omega^\mathrm{Obs}_\mathrm{M} \sim 0.3$ and $\Omega_\mathrm{Tot} \equiv 1$. 

Examining our 60 simulated cosmological models quantitatively, 
we informally choose a set of `best' runs -- six $\Psi _{\mathrm{Lin}}$ 
models and six $\Psi _{\mathrm{MD}}$ models -- which provide very good 
SNe data fits, while simultaneously producing good cosmological 
parameters. A truly optimized search over the $(\Psi _{0},z_\mathrm{init})$ 
parameter space is not really called for at this toy-model stage 
of our formalism; but a quick effort at optimization reveals an extensive 
`trench' in parameter space for the $\Psi _{\mathrm{MD}}$ runs with 
quite low (and highly degenerate) $\chi ^{2} _{\mathrm{Fit}}$ values, 
yet offering a wide variation to choose from regarding their output 
cosmological parameters. So for illustration purposes, we arbitrarily 
select one low-$\chi ^{2} _{\mathrm{Fit}}$ case, 
$(\Psi _{0},z_\mathrm{init}) = (0.768,14)$, as a so-called 
``semi-optimized" run, giving us a total of thirteen ``best runs" 
for more in-depth study. 

\begin{figure}
\begin{center}
\includegraphics[scale=1.0]{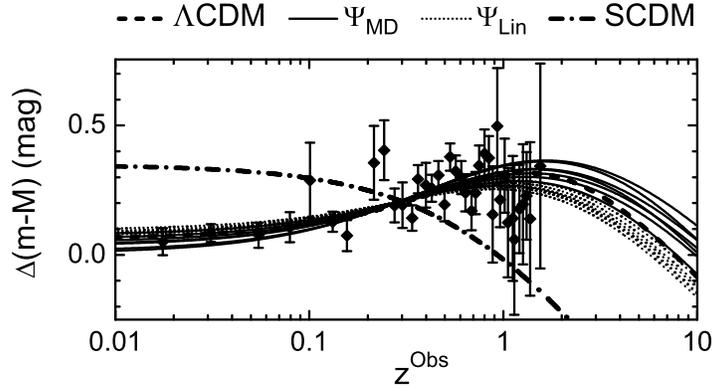}
\end{center}
\caption{Residual Hubble diagrams for our thirteen 
`best' models, shown along with the best-fit flat SCDM 
and Concordance $\Lambda{\mathrm{CDM}}$ 
($\Omega _{\Lambda} = 0.713$) cosmologies, and the binned 
and averaged SCP Union SNe data. From highest to lowest 
(at $z^\mathrm{Obs} \simeq 10$), the plotted 
$\Psi _{\mathrm{MD}} (t)$ curves have the parameters: 
$(\Psi _{0},z_\mathrm{init}) = (0.96,5)$, $(0.92,5)$, 
$(0.85,5)$, $(0.78,10)$, $(0.78,15)$, $(0.768,14)$, 
and $(0.78,25)$. From highest to lowest 
(at $z^\mathrm{Obs} \simeq 10$), the plotted 
$\Psi _{\mathrm{Lin}} (t)$ curves have the 
parameters: $(\Psi _{0},z_\mathrm{init}) = (1.0,10)$, 
$(1.0,15)$, $(0.96,15)$, $(1.0,25)$, $(0.96,25)$, 
and $(0.92,25)$. The individually-optimized 
$H^\mathrm{Obs}_{0}$ values for each model are depicted 
as differing vertical offsets for the curves.}
\label{FigBestCosmSims}
\end{figure}

Residual Hubble diagrams of these best runs are shown in 
Figure~\ref{FigBestCosmSims}. These thirteen cosmological models clearly 
produce good Hubble curves, being visually almost indistinguishable 
from one another (and from the now best-fit, $\Omega _{\Lambda} = 0.713$, 
Concordance $\Lambda$CDM model) in the SN-data-rich region of 
$z^{\mathrm{Obs}} \sim 0.1 - 1$. The causal backreaction formalism has 
therefore succeeded in reproducing the apparent cosmic acceleration 
as it is actually measured, via SNe standard candles.

\placefigure{FigBestCosmSims}

More quantitatively, the comprehensive output data from these thirteen 
best runs (and from best-fit $\Lambda$CDM and SCDM, for comparison) are 
given in Table~\ref{TableSimRunsCosParamsABBREV}.

\placetable{TableSimRunsCosParamsABBREV}

Summarizing the results: these runs fit the SNe data essentially 
as well as $\Lambda$CDM in terms of $\chi^{2}_{\mathrm{Fit}}$, and are 
comparable in $P_{\mathrm{Fit}}$. The mid-range values of $I_{0}$ 
indicate a strong enough backreaction effect to explain the apparent 
acceleration, without being so large (i.e., $I_{0} \sim 1$) as to 
indicate a substantial breakdown of our approximation of gravitational 
linearity in simply summing together individual metric perturbations 
to produce the total overall perturbation. 

Crucially, these models all solve the Age Problem 
without a Dark-Energy-induced acceleration, with 
$t^\mathrm{Obs}_{0} \sim 13.2 - 14.5$ GYr; and they also achieve 
(pre-perturbation) spatial flatness to within a reasonable degree 
of uncertainty. Specifically, the $\Psi _{\mathrm{Lin}}$ runs yield 
$\Omega^\mathrm{FRW}_\mathrm{M} \sim 0.90 - 1.04$, very close to unity; 
and while the $\Omega^\mathrm{FRW}_\mathrm{M}$ values for several of the 
$\Psi _{\mathrm{MD}}$ runs are somewhat high, lowering the (pre-adopted) 
normalization of $\Omega^\mathrm{Obs}_\mathrm{M} \equiv 0.27$ to the 
(also reasonable) value of $\Omega^\mathrm{Obs}_\mathrm{M} \equiv 0.24$ 
(and dropping the worst $\Psi _{\mathrm{MD}}$ run) moves them down to 
$\Omega^\mathrm{FRW}_\mathrm{M} \sim 0.88 - 1.25$. These $\sim$dozen runs 
also succeed fairly well in bracketing the SNe-best-fit $\Lambda$CDM value 
(for no radiation) of 
$l^\mathrm{Obs}_{\mathrm{A}, \Lambda \mathrm{CDM}} = 285.4$ for the CMB, 
performing vastly better than anything achievable with a matter-only 
{\it open} (i.e., $\Omega _\mathrm{M} \sim 0.3$) universe. 

Possessing very negative effective equation of state values, 
$w^\mathrm{Obs}_{0} < -0.7$, these models clearly produce a 
strong amount of apparent acceleration, despite the absence of 
a negative-pressure cosmic component; and we note that since 
the precise cosmological parameters obtained from SNe fits are 
highly model-dependent \citep[e.g.,][]{CattVisserCosmographSNeFits}, 
it is therefore unnecessary for alternative-cosmology models 
to {\it exactly} reproduce the $w^\mathrm{Obs}_{0}$ result from 
Concordance $\Lambda$CDM (e.g., $w^\mathrm{Obs}_{0} = -0.713$), 
but merely to generate an `accelerative' value of $w^\mathrm{Obs}_{0}$ 
strong enough to provide a sufficiently good fit to the supernova 
data (as has indeed been done here). Taken together, these results 
of a sufficiently-strong apparent acceleration, quantitatively good 
SNe fits, and good cosmological parameters all indicate that we have 
found a selection of causal backreaction models that successfully 
achieves the basic goals of an alternative cosmic concordance, 
without the use of any form of Dark Energy. 

Lastly, consider the observed jerk parameter, $j^\mathrm{Obs}_{0}$. 
Since flat $\Lambda$CDM cosmologies possessing only dust and a 
Cosmological Constant have $j(z) = j_{0} = 1$ for all time (for 
{\it any} choice of $\Omega _{\Lambda} = 1 - \Omega _\mathrm{M}$), 
searching for deviations from $j_{0} = 1$ represents a 
fundamental test of the FLRW-$\Lambda$CDM paradigm, itself. 
Coming from the third-order series term for 
$d_{\mathrm{L}}(z^\mathrm{Obs})$, $j^\mathrm{Obs}_{0}$ 
is not yet observationally well-constrained, and so can be 
considered a {\it testable prediction} of our formalism. 

From Table~\ref{TableSimRunsCosParamsABBREV} we have 
$j^\mathrm{Obs}_{0} \sim 2.5 - 5.5$ for our thirteen best runs, 
a range that can be further narrowed down to 
$j^\mathrm{Obs}_{0} \sim 2.6 - 3.8$ by restricting consideration 
to the six of those runs which are most consistent with all other 
cosmological parameter constraints (specifically: 
$\Psi _{\mathrm{Lin}}$ with $(\Psi _{0},z_\mathrm{init}) = (1.0,25)$, 
$(1.0,15)$, and $(0.96,25)$; and $\Psi _{\mathrm{MD}}$ with 
$(\Psi _{0},z_\mathrm{init}) = (0.78,10)$, $(0.85,5)$, and 
$(0.92,5)$). These jerk parameter values are clearly 
in excess of the $\Lambda$CDM prediction of $j_{0} = 1$, making 
a comparison with observational results imperative. Though the 
great difficulty of precisely measuring this parameter has led 
to a wildly-varying range of estimates, one potentially reliable 
estimate has been adapted in \citet{BochnerAccelPaperI} from the 
detailed results (D. Rubin, 2010, private communication) of the 
(SNe+CMB+BAO) analysis done with the SCP Union2 SNe compilation data 
\citep{AmanRubinSCPunion2}, using statistical and systematic errors, 
which in terms of the jerk parameter yields the result: 
$j^\mathrm{Obs}_{0} \pm \Delta j^\mathrm{Obs}_{0} \simeq 
2.129 \pm 1.292$; i.e., $j^\mathrm{Obs}_{0} \sim 0.8 - 3.4$. 
While still consistent with Concordance $\Lambda$CDM, it is 
focused predominantly on the high side, as generally predicted 
by our formalism (if not {\it quite} so high as several of our 
models predict); an observationally interesting situation, 
being not yet conclusive in either direction. 

One important caveat, however, is that we cannot conclude with 
certainty -- based only upon the results in this paper -- that 
high jerk parameters are a robust feature of our formalism. 
Too many theoretical uncertainties exist within this toy-model 
approach to be sure, just yet, that this is an `iron-clad' 
prediction of causal backreaction. This includes our neglect 
here of both recursive and gravitational nonlinearities; as 
well as the oversimplified nature of these $\Psi (t)$ clumping 
evolution models, which neglect astrophysical processes such as 
feedback from star formation and the shock-heating of baryons 
\citep[e.g.,][]{CenOstrikerShockedOb}. All such effects 
(except perhaps gravitational nonlinearities) should result in 
the damping of the apparent acceleration, especially at later 
times, resulting in $j^\mathrm{Obs}_{0}$ values that are reduced 
compared to those seen in the output results of the naive models 
presented here. Thus the theoretical situation is as uncertain 
as the observational situation is with regard to this key 
cosmological parameter, and more theoretical development of the 
causal backreaction formalism needs to be done before this paradigm 
can be conclusively tested by measurements of $j^\mathrm{Obs}_{0}$. 
It is possible, in any case, that other means may be found to test 
it; for example, the fact that causal backreaction does not occur 
in completely smoothly-distributed fashion, but is instead most 
concentrated near virializing masses, may lead to interesting 
feedback behaviors affecting the formation of stars, galaxies and 
galaxy clusters. Conceivably, this may have some useful application 
to issues such as galaxy downsizing \citep{CowieDownsizing}, 
the cuspy CDM halo problem, and the possible dearth of dwarf 
satellite galaxies \citep[e.g.,][]{PrimackCuspy}; but such 
connections are speculative, and finding a variety of concrete 
methods for testing causal backreaction remains of 
great importance for this paradigm.

\section{\label{SecConcludeAndSummary}SUMMARY AND CONCLUSIONS}

Our starting point in this paper has been the search for an 
alternative to Dark Energy for attaining a Cosmic Concordance, 
without resorting to the added complexity of modifications to gravity, 
or the non-Copernican `specialness' implied by a local cosmic void. 
The problems with Dark Energy are well known, including the magnitude 
and coincidence fine-tuning problems for a pure-$\Lambda$ Cosmological 
Constant, and the stability problem for a more dynamical form of Dark 
Energy possessing the ability to cluster spatially. The latter (DDE) 
case requires the ad-hoc addition of some form of new pressure term 
to support it against collapse, given the locally attractive nature 
of the negative pressure required by the DDE to power the 
cosmic acceleration. Such a new pressure term, if adiabatic 
(e.g., degeneracy pressure), would represent a form of {\it positive} 
pressure contributing to a cosmic {\it deceleration} that partially 
or totally nullifies the acceleration meant to come from the DDE; and 
if non-adiabatic (due to some new effects from the DDE Lagrangian) 
would invalidate the Dark Energy as a perfect fluid, thus calling 
into question the usual cosmic implications of $w ^{\mathrm{DE}} < 0$ 
in the FLRW acceleration equation. 

Emphasizing the importance of the fact that `negative pressure' 
is locally attractive in character -- rather than repulsive, as it is 
often regarded and popularly described -- and that normal gravitational 
attraction therefore represents (at least in a non-technical sense) a 
form of negative pressure, we have therefore made a virtue of necessity 
by recruiting cosmological structure formation itself, based upon 
ordinary gravitational forces, as the driver of the observed (possibly 
apparent) acceleration. This approach, known generally in the literature 
as ``backreaction'', removes all of the aforementioned problems by 
eliminating the need for any Dark Energy species, while solving the 
Coincidence Problem by naturally linking the onset of cosmic 
acceleration to the emergence of cosmic structure. This linkage 
inevitably leads to the creation of observers (such as ourselves) 
just in time to see this `coincidence'. 

Noting that different varieties of backreaction proposed by other 
authors have been largely unsuccessful in their attempts to account 
for the observed acceleration, and that powerful arguments have been 
advanced which claim that backreaction as a paradigm {\it cannot} be 
made strong enough to succeed at this task, we have explained why 
each of these ``no-go'' arguments are functionally invalid, due to 
a number of astrophysically inappropriate approximations. These 
include the dropping of (squared) vorticity and velocity dispersion 
terms in backreaction calculations; the unrealistic boundary 
conditions and isolation of overdensities in Swiss-Cheese models; 
and the neglect of Newtonian-level perturbations, due to well-known 
results from an influential backreaction formalism (and from 
perturbation theory in general) which adopt an inappropriate 
`slow-motion' approximation, and thus lack essential effects 
due to the causal propagation of gravitational information 
from developing inhomogeneities (``causal updating''). 

Informed by the consequential but flawed assumptions of such 
no-go arguments, we have developed a phenomenological formalism 
of ``causal backreaction'', in which the apparent acceleration of 
the universe is the result of the universal phase transition from 
smoothness to clumpiness, and is achieved through the combined 
perturbative contributions of all virializing, self-stabilizing 
cosmic structures to the metric of a typical cosmological observer. 
Averaging over location and direction, the evolution of this 
``smoothly-inhomogeneous'' cosmology is determined by a single 
input function, $\Psi (t)$, chosen from a selection of clumping 
evolution models which are specified heuristically but guided 
by observed astrophysical data. 

Using a set of such clumping evolution models with physically-motivated 
input parameters, we have conducted 60 simulation runs, finding a 
$\sim$dozen-plus solutions that fit the SCP Union Compilation Type Ia 
supernova data essentially as well as the best-fit $\Lambda$CDM model. 
Furthermore, about half of these models give good values (within 
reasonable theoretical and observational uncertainties) for all other 
calculated cosmological parameters studied here, such as: the age of 
the universe (thus solving the Cosmic Age Problem); the matter density 
as a fraction of the unperturbed FRW critical density (thereby achieving 
effective spatial flatness with matter alone); the deceleration parameter 
(therefore achieving a sufficiently-strong apparent acceleration); and 
the characteristic angular scale of the CMB acoustic peaks (thus matching 
observations from the early, pre-structure-formation cosmic epoch). All 
of these goals are achieved without altering the measured physical 
density of cosmic matter, and without the introduction of any form of 
Dark Energy, thus forming the basis of an alternative concordance for 
a matter-only universe. 

Looking for testable ways to distinguish our causal backreaction formalism 
from Cosmological Constant $\Lambda$CDM (and from other slowly-evolving 
forms of Dark Energy not too far from $\Lambda$), we have seen that while 
the jerk parameter must always obey the condition $j^\mathrm{Obs}_{0} = 1$ 
for flat $\Lambda$CDM, the results of those of our causal backreaction models 
which successfully fit the supernova data instead predict $j^\mathrm{Obs}_{0}$ 
values strongly in excess of unity, a result which has some support from the 
observational data. Nevertheless, we note these caveats: $\Lambda$CDM still 
remains consistent at the $1 \sigma$ level with most observations; the 
definitive observational constraints on $j^\mathrm{Obs}_{0}$ remain 
extremely weak, in any case; and perhaps most importantly, theoretical 
uncertainties for the highly simplified causal backreaction formalism 
introduced here has likely led to overestimates in our predicted 
values of the jerk parameter, a potential problem which must be 
remedied through the further theoretical development of the model. 

Expressing the conclusions of this paper in one sentence: if someone were 
to ask, ``What is the force behind the cosmic acceleration?'', the answer 
we would give is that it is not a `force' at all; rather, the motivating 
factor is the total, summed effect of the Newtonian tails of individual 
metric perturbations, produced by the virialization of innumerable 
self-stabilizing structures filling the universe, with these influences 
propagating causally towards all observers from the extreme edges of 
their observable cosmic horizons.

\acknowledgments

I am grateful to Jacob Bekenstein, Krzysztof Bolejko, 
Varoujan Gorjian, Wayne Hu, Marek Kowalski, 
Edvard M\"{o}rtsell, David Rubin, Ran Sivron, 
David Wiltshire, and Ned Wright for brief but helpful 
communications; and I am especially grateful to 
Arthur Lue for several helpful and clarifying 
discussions.


\newpage


\begin{table}
\begin{center}
\caption{Cosmological Parameters Derived from 
our `Best' Simulation Runs 
\label{TableSimRunsCosParamsABBREV}}
\resizebox{16.5cm}{!} 
{
\begin{tabular}{cccccccccccc}
\tableline
\tableline
($\Psi _{0},z_\mathrm{init}$) &
$\chi^{2}_{\mathrm{Fit}}$ & 
$P_{\mathrm{Fit}}$ \tablenotemark{a} & 
$I_{0}$ & 
$z^\mathrm{Obs}$ \tablenotemark{b} & 
$H^\mathrm{Obs}_{0}$ \tablenotemark{c} & 
$H^\mathrm{FRW}_{0}$ \tablenotemark{d} & 
$t^\mathrm{Obs}_{0}$ \tablenotemark{e} & 
$\Omega^\mathrm{FRW}_\mathrm{M}$ \tablenotemark{f} & 
$w^\mathrm{Obs}_{0}$ & 
$j^\mathrm{Obs}_{0}$ & 
$l^\mathrm{Obs}_{\mathrm{A}}$ 
\\
\tableline
\multicolumn{12}{c}{{\it $\Psi _{\mathrm{Lin}}$ Clumping Model Runs}}\\
\tableline
(1.0,25) & 312.1 & 0.362 & 0.57 & 1.12 & 69.95 & 40.68 & 13.56 & 
1.037 & -0.817 & 3.45 & 284.2 \\
(1.0,15) & 313.3 & 0.344 & 0.55 & 1.12 & 69.64 & 41.63 & 13.45 & 
0.971 & -0.784 & 3.11 & 287.6 \\
(1.0,10) & 315.4 & 0.314 & 0.52 & 1.12 & 69.27 & 42.80 & 13.30 & 
0.897 & -0.746 & 2.76 & 291.3 \\
(0.96,25) & 314.8 & 0.323 & 0.55 & 1.11 & 69.38 & 41.57 & 13.38 & 
0.968 & -0.759 & 2.92 & 285.9 \\
(0.96,15) & 316.6 & 0.297 & 0.52 & 1.11 & 69.11 & 42.46 & 13.28 & 
0.911 & -0.732 & 2.67 & 289.1 \\
(0.92,25) & 319.0 & 0.265 & 0.53 & 1.11 & 68.85 & 42.44 & 13.21 & 
0.905 & -0.706 & 2.49 & 287.6 \\
\tableline
\multicolumn{12}{c}{{\it $\Psi _{\mathrm{MD}}$ Clumping Model Runs}}\\
\tableline
(0.78,10) & 312.1 & 0.362 & 0.63 & 1.12 & 69.96 & 38.25 & 13.88 & 
1.204 & -0.801 & 3.15 & 277.5 \\
(0.78,15) & 312.2 & 0.360 & 0.68 & 1.12 & 70.84 & 36.14 & 14.17 & 
1.409 & -0.895 & 4.17 & 270.5 \\
(0.78,25) & 316.8 & 0.295 & 0.72 & 1.12 & 71.80 & 34.22 & 14.46 & 
1.642 & -1.001 & 5.51 & 263.7 \\
(0.85,5) & 313.9 & 0.336 & 0.56 & 1.13 & 69.48 & 41.21 & 13.60 & 
0.991 & -0.747 & 2.59 & 288.7 \\
(0.92,5) & 312.1 & 0.363 & 0.60 & 1.14 & 70.71 & 39.41 & 14.00 & 
1.144 & -0.871 & 3.75 & 285.6 \\
(0.96,5) & 315.5 & 0.313 & 0.63 & 1.14 & 71.52 & 38.36 & 14.25 & 
1.248 & -0.954 & 4.67 & 283.8 \\
\tableline
\multicolumn{12}{c}{{\it Semi-Optimized 
$\Psi _{\mathrm{MD}}$ Clumping Model Run}}\\
\tableline
(0.768,14) & 311.7 & 0.369 & 0.66 & 1.12 & 70.37 & 36.91 & 14.03 & 
1.324 & -0.845 & 3.63 & 272.4 \\
\tableline
\multicolumn{12}{c}{{\it Comparison Values from 
Best-Fit flat $\Lambda{\mathrm{CDM}}$ Model 
$(\Omega _{\Lambda} = 0.713 = 1 - \Omega _\mathrm{M})$}} 
\\
\tableline
\nodata & 311.9 & 0.380 & \nodata & 1.0 & 69.96 & 69.96 & 13.64 & 
0.287 & -0.713 & 1.0 & 285.4 \\
\tableline
\multicolumn{12}{c}{{\it Comparison Values from 
Best-Fit flat SCDM Model 
$(\Omega _{\Lambda} = 0$, $\Omega _\mathrm{M} = 1)$}}\\
\tableline
\nodata & 608.2 & 3.4E-22 & \nodata & 1.0 & 61.35 & 61.35 & 10.62 & 
1.0 & 0.0 & 1.0 & 287.3 \\
\tableline
\end{tabular}
}
\tablenotetext{a}{$P_{\mathrm{Fit}}$ values are calculated using 
$N_{\mathrm{DoF}} = 304$ for the $\Psi _{\mathrm{Lin}}$ and 
$\Psi _{\mathrm{MD}}$ models, $N_{\mathrm{DoF}} = 305$ 
for flat $\Lambda{\mathrm{CDM}}$, and $N_{\mathrm{DoF}} = 306$ 
for flat SCDM.}
\tablenotetext{b}{Each $z^\mathrm{Obs}$ value quoted here 
corresponds to $z^\mathrm{FRW} \equiv 1$.}
\tablenotetext{c}{All $H_{0}$ values are given in 
$\mathrm{km} ~ \mathrm{s}^{-1} \mathrm{Mpc}^{-1}$.}
\tablenotetext{d}{Each $H^\mathrm{FRW}_{0}$ is computed relative 
to the corresponding optimized $H^\mathrm{Obs}_{0}$ value 
for that run.}
\tablenotetext{e}{$t^\mathrm{Obs}_{0}$ values are listed in GYr, 
and all quantities are computed assuming {\it no radiation} 
($\Omega _\mathrm{R} \equiv 0$).}
\tablenotetext{f}{All $\Omega^\mathrm{FRW}_\mathrm{M}$ values given 
for the $\Psi _{\mathrm{Lin}}$ and $\Psi _{\mathrm{MD}}$ models 
are normalized to $\Omega^\mathrm{Obs}_\mathrm{M} \equiv 0.27$.}
\end{center}
\end{table}

\end{document}